\documentclass[seceq]{ptptex}

\usepackage{wrapft}

\title{Canonical partition
function for anomalous systems\\ described by the
$\kappa$-entropy}

\author{Antonio M.
\textsc{Scarfone}$^{1,}$\footnote{antonio.scarfone@polito.it}
and Tatsuaki
\textsc{Wada}$^{2,}$\footnote{wada@ee.ibaraki.ac.jp} }

\inst{$^1$Istituto
Nazionale di Fisica della Materia (INFM) and Dipartimento di Fisica\\
Unit\'a del Politecnico di Torino, Corso Duca degli Abruzzi
24,
I-10129 Torino, Italy\\
\vspace{3mm} $^2$Department of Electrical and Electronic
Engineering,\\ Ibaraki University, Hitachi, Ibaraki,
316-8511, Japan }

\abst{Starting from the $\kappa$-distribution function,
obtained by applying the maximal entropy principle to the
$\kappa$-entropy [G. Kaniadakis, Phys.\ Rev.\ E\
\textbf{66} (2002), 056125], we derive the expression of
the canonical $\kappa$-partition function and discuss its
main properties. It is shown that all important
macroscopical quantities of the system can be expressed
employing only the $\kappa$-partition function. The
relationship between the associated $\kappa$-free energy
and the $\kappa$-entropy is also discussed. }

\begin{document}

\maketitle

\section{Introduction}
Anomalous statistical systems which exhibit an asymptotic
power law behavior in the probability distribution function
(pdf) are ubiquitous in
nature.\cite{Abe}\tocite{Kaniadakis00} Although not in
equilibrium these distributions characterize a metastable
configuration in which the system remains for a very long
period of time compared to the typical time scales of its
underlying
microscopical dynamics.\\
The statistical properties of such systems can be
investigated trough the introduction of a generalization of
the Boltzmann-Gibbs (BG) entropy $S^{\rm
B}=-\sum_ip_{_i}\,\ln(p_{_i})$ with
$p\equiv\{p_{_i}\}_{i=1,\,\cdots,\,N}$ a discrete pdf,
(throughout this paper we adopt units within $k_{\rm B}=1$)
from which, by means of the maximal entropy principle, the
corresponding pdf can be derived.

Recently, in\cite{Kaniadakis1}\tocite{Kaniadakis3} it has
been proposed a generalized statistical mechanics based on
the deformed entropy
\begin{equation}
S_{_\kappa}=-\sum_ip_{_i}\,\ln_{_{\{\kappa\}}}(p_{_i}) \
,\label{kentropy}
\end{equation}
which preserves the structure of the BG
statistical mechanics.\\
Eq. (\ref{kentropy}) mimics the well-known classical
entropy by replacing the standard logarithm with the
generalized version
\begin{equation}
\ln_{_{\{\kappa\}}}(x)=\frac{x^\kappa-x^{-\kappa}}{2\,\kappa}
\ ,\label{klog}
\end{equation}
where $-1<\kappa<1$,
$\ln_{_{\{\kappa\}}}(x)=\ln_{_{\{-\kappa\}}}(x)$ and
$\ln_{_{\{\kappa\}}}(x)=-\ln_{_{\{\kappa\}}}(1/x)$. Eq.
(\ref{klog}) reduces to the classical logarithm in the
$\kappa\to0$ limit: $\ln_{_{\{0\}}}(x)=\ln(x)$, as well as,
in the same limit, Eq. (\ref{kentropy}) reduces to the BG
entropy.\\
The $\kappa$-entropy (\ref{kentropy}) has many properties
of the standard entropy like positivity, continuity,
symmetry, expansibility, decisivity, maximality and
concavity\cite{Scarfone1}. Moreover, in\cite{Scarfone2} it
has been shown that the $\kappa$-entropy is also Lesche
stable, an important property that must be fulfilled in
order to represent a well defined physical observable.\\
The maximization of Eq. (\ref{kentropy}) under the
constraints on the normalization and on the mean energy
\begin{equation}
\sum_ip_{_i}=1 \ ,\hspace{20mm}\sum_iE_{_i}\,p_{_i}=U \
,\label{const}
\end{equation}
leads to the following $\kappa$-distribution
function\cite{Kaniadakis3}
\begin{equation}
p_{_i}=\alpha\,\exp_{_{\{\kappa\}}}
\left(-{1\over\lambda}\left(\gamma+\beta\,E_{_i}\right)\right)
\ ,\label{distr}
\end{equation}
where $\gamma$ and $\beta$ are the Lagrange multipliers
associates to the constraints (\ref{const}). The deformed
exponential $\exp_{_{\{\kappa\}}}(x)$, the inverse function
of $\ln_{_{\{\kappa\}}}(x)$, is defined as
\begin{equation}
\exp_{_{\{\kappa\}}}(x)=\left(\kappa\,x+\sqrt{1+\kappa^2\,x^2}\right)^{1/\kappa}
\ ,\label{kexp}
\end{equation}
and reduces to the standard exponential in the $\kappa\to0$
limit: $\exp_{_{\{0\}}}(x)=\exp(x)$.\\
The constants $\lambda$ and $\alpha$ in Eq. (\ref{distr})
are given by
\begin{eqnarray}
\lambda=\sqrt{1-\kappa^2} \ ,\hspace{20mm}
\alpha=\left(\frac{1-\kappa}{1+\kappa}\right)^{1/2\,\kappa}
\ ,
\end{eqnarray}
respectively, and are related each other through the
relation $\ln_{_{\{\kappa\}}}(\alpha)=-1/\lambda$.

In\cite{Kaniadakis3,Scarfone3} it has been shown that,
starting from the deformed logarithm (\ref{klog}) and the
deformed exponential (\ref{kexp}), the $\kappa$-algebra can
be developed in a way that many algebraic properties of the
standard logarithm and exponential can be reproduced in the
deformed version. For instance, through the definition of
the $\kappa$-sum
$x\oplus\mbox{\raisebox{3mm}{\hspace{-3mm}$\scriptscriptstyle
{\kappa}$}} \hspace{2mm}y$, given by
\begin{equation}
x\oplus\mbox{\raisebox{3mm}{\hspace{-3mm}$\scriptscriptstyle
{\kappa}$}}
\hspace{2mm}y=x\,\sqrt{1+\kappa^2\,y^2}+y\,\sqrt{1+\kappa^2\,x^2}
\ ,\label{ksum}
\end{equation}
it is easy to verify the following useful relation
\begin{equation}
\exp_{_{\{\kappa\}}}\left(x\oplus\mbox{\raisebox{3mm}{\hspace{-3mm}$\scriptscriptstyle
{\kappa}$}}
\hspace{2mm}y\right)=\exp_{_{\{\kappa\}}}(x)\cdot\exp_{_{\{\kappa\}}}(x)
\ .\label{expsum}
\end{equation}
In\cite{Kaniadakis3,Kaniadakis4} it has been shown that the
$\kappa$-statistical mechanics emerges within the special
relativity. The physical mechanism introducing the
$\kappa$-deformation is originated from the Lorentz
transformations and the $\kappa\to0$ limit, reproducing the
ordinary statistical mechanics, corresponds to the
classical limit $c\to\infty$.

The purpose of this work is to introduce the generalized
canonical partition function $Z_{_{\kappa}}$ in the
framework of the $\kappa$-deformed statistical mechanics
and to show that all the relevant relations, valid in the
BG theory, still hold in the deformed version.\\

We start by recalling the useful relations. Firstly, the
$\kappa$-logarithm fulfills the functional-differential
equation\cite{Kaniadakis3}\tocite{Scarfone1}
\begin{equation}
\frac{d}{d\,x}\left[x\,\Lambda(x)\right]
=\lambda\,\Lambda\left({x\over\alpha}\right) \
,\label{difeq}
\end{equation}
with the boundary conditions $\Lambda(1)=0$ and
$(d/d\,x)\,\Lambda(x)\Big|_{x=1}=1$.\\
Another solution of Eq. (\ref{difeq}), which follows from
the boundary conditions $\Lambda(1)=1$ and
$(d/d\,x)\Lambda(x)\Big|_{x=1}=0$, is given
by\cite{Scarfone-Wada}
\begin{equation}
u_{_{\{\kappa\}}}(x)=\frac{x^\kappa+x^{-\kappa}}{2} \
.\label{ku}
\end{equation}
It fulfills the relations
$u_{_{\{\kappa\}}}(x)=u_{_{\{-\kappa\}}}(x)$,
$u_{_{\{\kappa\}}}(x)=
u_{_{\{\kappa\}}}(1/x)$ and $u_{_{\{\kappa\}}}(\alpha)=1/\lambda$.\\
From the definitions (\ref{klog}) and (\ref{ku}) we obtain
the useful equations
\begin{eqnarray}
\ln_{_{\{\kappa\}}}(x\,y)&=&u_{_{\{\kappa\}}}(x
)\,\ln_{_{\{\kappa\}}}(y)+\ln_{_{\{\kappa\}}}(x
)\,u_{_{\{\kappa\}}}(y) \ ,\label{logsum}\\
u_{_{\{\kappa\}}}(x\,y)&=&u_{_{\{\kappa\}}}(x
)\,u_{_{\{\kappa\}}}(y)+\kappa^2\,
\ln_{_{\{\kappa\}}}(x)\,\ln_{_{\{\kappa\}}}(y) \
,\label{usum}
\end{eqnarray}
showing a deep link between both the functions
$\ln_{_{\{\kappa\}}}(x)$
and $u_{_{\{\kappa\}}}(x)$.\\
Starting from Eq. (\ref{ku}), in analogy with Eq.
(\ref{kentropy}), we introduce the function
\begin{equation}
{\cal I}_{_\kappa}=\sum_ip_{_i}\,u_{_{\{\kappa\}}}(p_{_i})
\ ,\label{ki}
\end{equation}
which can also be defined as the mean value of
$u_{_{\{\kappa\}}}(x)$ according to the relation ${\cal
I}_{_\kappa}=\langle u_{_{\{\kappa\}}}(p)\rangle$. The
$\kappa$-entropy
$S_{_\kappa}=\langle\ln_{_{\{\kappa\}}}(p)\rangle$ as well,
can be
defined as the mean value of $\ln_{_{\{\kappa\}}}(x)$.\\
It is worth to observe that, by using the definitions
(\ref{kentropy}) and (\ref{ki}), we obtain from Eq.
(\ref{logsum}) the equation\cite{Scarfone-Wada}
\begin{equation}
S_{_\kappa}({\rm A}\cup{\rm B})={\cal I}_{_\kappa}({\rm
A})\,S_{_\kappa}({\rm B})+S_{_\kappa}({\rm A})\,{\cal
I}_{_\kappa}({\rm B}) \ ,\label{sum}
\end{equation}
stating the additivity rule of the $\kappa$-entropy for two
statistically independent systems A and B, in the sense of
$p^{{\rm A}\cup{\rm B}}=p^{\rm A}\cdot p^{\rm B}$. In the
$\kappa\to0$ limit $u_{_{\{0\}}}(p_{_i})=1$ and ${\cal
I}_{_0}(p)=\sum_i p_{_i}=1$ so that Eq. (\ref{sum})
recovers the additivity rule of the BG entropy.

Returning to the $\kappa$-distribution, we pose
\begin{equation}
x_{_i}=\gamma+\beta\,E_{_i} \ ,
\end{equation}
so that Eq. (\ref{distr}) can be written in
\begin{eqnarray}
\nonumber p_{_i}&=&\alpha\,\exp_{_{\{\kappa\}}}
\left(-{x_{_i}\over\lambda}\right)=\exp_{_{\{\kappa\}}}\left(-{1\over\lambda}\right)\cdot\exp_{_{\{\kappa\}}}
\left(-{x_{_i}\over\lambda}\right)\\
&=&\exp_{_{\{\kappa\}}}\left[\left(-{1\over\lambda}\right)
\oplus\mbox{\raisebox{3mm}{\hspace{-3mm}$\scriptscriptstyle
{\kappa}$}}
\hspace{1mm}\left(-{x_{_i}\over\lambda}\right)\right] \
,\label{distr1}
\end{eqnarray}
where Eq. (\ref{expsum}) has been employed. By using the
definition of the $\kappa$-sum the argument in Eq.
(\ref{distr1}) can be rewritten in
\begin{equation}
\left(-{1\over\lambda}\right)\oplus\mbox{\raisebox{3mm}{\hspace{-3mm}$\scriptscriptstyle
{\kappa}$}}
\hspace{1mm}\left(-{x_{_i}\over\lambda}\right)=-{1\over\lambda}\,\left({x_{_i}\over\lambda}
+\sqrt{1+{\kappa^2\over\lambda^2}\,x_{_i}^2}\right) \ ,
\end{equation}
and Eq. (\ref{distr1}) becomes
\begin{equation}
p_{_i}=\exp_{_{\{\kappa\}}}\left[-{1\over\lambda}\,\left({x_{_i}\over\lambda}
+\sqrt{1+{\kappa^2\over\lambda^2}\,x_{_i}^2}\right)\right]
\ .\label{distr2}
\end{equation}
On the other hand, by using Eq. (\ref{usum}) with
$x=\alpha$ and $y=\exp_{_{\{\kappa\}}}(-x_{_i}/\lambda)$,
and taking into account the relation
\begin{equation}
u_{_{\{\kappa\}}}\Bigg(\exp_{_{\{\kappa\}}}\left(-{x_{_i}\over\lambda}\right)\Bigg)
=\sqrt{1+{\kappa^2\over\lambda^2}\,x_{_i}^2} \ ,
\end{equation}
which follows from the definitions (\ref{kexp}) and
(\ref{ku}), we obtain
\begin{equation}
u_{_{\{\kappa\}}}\left(p_{_i}\right)={1\over\lambda}\,\left({\kappa^2\over\lambda}\,x_{_i}
+\sqrt{1+{\kappa^2\over\lambda^2}\,x_{_i}^2}\right) \ ,
\end{equation}
so that Eq. (\ref{distr2}) can be easily written as
\begin{equation}
p_{_i}=\exp_{_{\{\kappa\}}}\,
\left(-u_{_{\{\kappa\}}}\left(p_{_i}\right)-\gamma-\beta\,E_{_i}\right)
\ ,\label{distr3}
\end{equation}
which is an alternative but equivalent expression of the
$\kappa$-distribution (\ref{distr}).

\section{Canonical partition function}

We recall that in the classical theory the canonical
partition function $Z$ is an important quantity that
encodes the statistical properties of a system. From the BG
distribution
\begin{equation}
p_{_i}=e^{-1-\gamma}\,e^{-\beta\,E_{_i}}={1\over
Z}\,e^{-\beta\,E_{_i}} \ ,\label{clasdist}
\end{equation}
it follows that the partition function can be introduced as
\begin{equation}
\ln (Z)=1+\gamma \ .\label{clasz}
\end{equation}
It depends firstly on the Lagrange multiplier $\beta$ and
secondly on the microstate energies $E_{_i}$ which are
determined by other macroscopical quantities like the
volume or the number of particles. Remarkably, most of the
thermodynamical functions of the system can be expressed in
terms of the partition function or its derivatives. For
instance, the entropy $S^{\rm BG}=\ln(Z)+\beta\,U$, the
total energy $U=-d\,\ln(Z)/d\,\beta$ and the free energy
$F=-\ln(Z)/\beta$.\\
We observe that, by using the expression (\ref{distr3}),
from the definition of the $\kappa$-entropy we obtain the
relation
\begin{eqnarray}
\nonumber
S_{_\kappa}&=&-\sum_ip_{_i}\,\ln_{_{\{\kappa\}}}\left(p_{_i}\right)=
\sum_ip_{_i}\,\Big(u_{_\kappa}\left(p_{_i}\right)
+\gamma+\beta\,E_{_i}\Big)\\
&=&{\cal I}_{_{\kappa}}+\gamma+\beta\,U \ ,\label{ent1}
\end{eqnarray}
which reminds us the classical expression
$S=1+\gamma+\beta\,U$, and it is recovered in the
$\kappa\to0$ limit.

According to Eq. (\ref{ent1}), taking into account its
limit for $\kappa\to0$ and the classical relationship
between the BG entropy and the canonical partition function
$Z$, we are guided to define the canonical
$\kappa$-partition function $Z_{_\kappa}$ through
\begin{equation}
\ln_{_{\{\kappa\}}}\left(Z_{_\kappa}\right)={\cal
I}_{_{\kappa}}+\gamma \ .\label{z}
\end{equation}
Eq. (\ref{z}) reduces to the definition (\ref{clasz}) in
the $\kappa\to0$ limit.

In order to verify the consistency of the definition
(\ref{z}), we derive its main properties in the framework
of the $\kappa$-statistical mechanics.\\
Firstly, it is trivial to verify that the entropy
(\ref{ent1}) becomes
\begin{equation}
S_{_\kappa}=\ln_{_{\{\kappa\}}}\left(Z_{_\kappa}\right)+\beta\,U
\ ,\label{ent2}
\end{equation}
which mimics the corresponding classical relationship.\\
Successively, we compute the derivative of the entropy
$S_{_\kappa}$ w.r.t the mean energy
\begin{eqnarray}
\nonumber \frac{d\,S_{_\kappa}}{d\,
U}&=&-\sum_i\frac{d}{d\,p_{_i}}\left[p_{_i}\,\ln_{_{\{\kappa\}}}
\left(p_{_i}\right)\right]\,\frac{d\,p_{_i}}{d\,U}=-\lambda\,\sum_i\ln_{_{\{\kappa\}}}
\left(\frac{p_{_i}}{\alpha}\right)\,\frac{d\,p_{_i}}{d\,U}\\
&=&\sum_i\left(\gamma+\beta\,E_{_i}\right)\,\frac{d\,p_{_i}}{d\,U}
\ ,
\end{eqnarray}
where we have taken into account Eqs. (\ref{distr}) and
(\ref{difeq}). Under the no-work condition $\sum_i
p_{_i}\,dE_{_i}=0$ (consequently $dU =\sum_i
E_{_i}\,dp_{_i}$), and recalling that $\sum_id\,p_{_i}=0$,
which follows from the normalization condition on the pdf,
we obtain
\begin{equation}
\frac{d\,S_{_\kappa}}{d\, U}=\beta \ .\label{ds}
\end{equation}
On the other hand, starting from Eq. (\ref{ent2}) we have
\begin{equation}
\frac{d\,S_{_\kappa}}{d\,U}=\frac{d}{d\,
U}\,\ln_{_{\{\kappa\}}}\left(Z_{_\kappa}\right)+U\,\frac{d\,
\beta}{d\,U}+\beta \ ,
\end{equation}
and by comparing this relation with Eq. (\ref{ds}) it
follows that, the following important property
\begin{equation}
\frac{d}{d\,\beta}\,\ln_{_{\{\kappa\}}}\left(Z_{_\kappa}\right)=-U
\ ,\label{zb}
\end{equation}
must hold.\\ Eq. (\ref{zb}) can be proved directly from the
definition (\ref{z}). In fact, by using Eq. (\ref{difeq})
we obtain
\begin{eqnarray}
\nonumber \frac{d}{d\,
\beta}\,\ln_{_{\{\kappa\}}}\left(Z_{_\kappa}\right)&=&\frac{d}{d\,
\beta}\left(I_{_{\kappa}}+\gamma\right)\\
\nonumber&=&\sum_i\frac{d}{d\,p_{_i}}\left[p_{_i}\,u_{_{\{\kappa\}}}
\left(p_{_i}\right)\right]\,\frac{d\,p_{_i}}{d\,\beta}+\frac{d\,\gamma}{d\,\beta}\\
&=&\lambda\sum_iu_{_{\{\kappa\}}}\left(\frac{p(x_{_i})}{\alpha}\right)\,
\frac{d\,p(x_{_i})}{d\,x_{_i}}\,\frac{d\,x_{_i}}{d\,\beta}
+\frac{d\,\gamma}{d\,\beta} \ ,\label{r1}
\end{eqnarray}
where $p(x_{_i})\equiv p_{_i}$. Using in Eq. (\ref{r1}) the
relation
\begin{equation}
p(x_{_i})=-\lambda\,u_{_{\{\kappa\}}}\left({p(x_{_i})\over\alpha}\right)\,\frac{d\,p(x_{_i})}{d\,x_{_i}}
\ ,
\end{equation}
we finally obtain
\begin{equation}
\frac{d}{d\,
\beta}\,\ln_{_{\{\kappa\}}}\left(Z_{_\kappa}\right)=
-\sum_ip_{_i}\left(\frac{d\,\gamma}{d\,\beta}+E_{_i}\right)
+\frac{d\,\gamma}{d\, \beta}=-U \ .
\end{equation}
Eq. (\ref{zb}) is the dual relation of Eq. (\ref{ent2}).
They state, from one hand that both $\beta $ and $U$ are
canonically conjugate variables, and on the other hand that
$S_{_\kappa}$ is a function of $U$ whereas $Z_{_\kappa}$ is
a function of $\beta$, like in the BG theory.

Accounting for the standard relationships between the free
energy and the partition function, we introduce the
$\kappa$-free energy
\begin{equation}
F_{_\kappa}=-{1\over\beta}\,\ln_{_{\{\kappa\}}}\left(Z_{_\kappa}\right)
\ .\label{f}
\end{equation}
It is trivial to observe that the definition (\ref{f}) can
be obtained by means of a Legendre transformation on the
mean energy
\begin{equation}
F_{_\kappa}=U-{d\,U\over d\,
S_{_\kappa}}\,S_{_\kappa}=U-{1\over\beta}\,S_{_\kappa} \ ,
\end{equation}
as it follows through Eqs. (\ref{ent2}) and (\ref{ds}).\\
Finally, we observe that the $\kappa$-free energy is a
function of $1/\beta$ as it follows from the relation
\begin{equation}
\frac{d\,F_{_\kappa}}{d(1/\beta)}=-S_{_\kappa} \ ,
\end{equation}
which mimics the classical relationship between the free
energy and the BG entropy.
\section{Mean value of an observable}

Another interesting relation involving the partition
function (\ref{z}) can be obtained by evaluating the
derivative of $\ln_{_{\{\kappa\}}}\left(Z_{_\kappa}\right)$
w.r.t the energy levels $E_{_i}$, for an equilibrium
(meta-equilibrium) state with $\beta\simeq const.$ We have
\begin{eqnarray}
\nonumber \frac{d}{d\,
E_{_i}}\,\ln_{_{\{\kappa\}}}\left(Z_{_\kappa}\right)&=&\frac{d}{d\,
E_{_i}}\,\left({\cal
I}_{_{\kappa}}+\gamma\right)\\
\nonumber&=&\sum_j\frac{d}{d\,p_{_j}}\left[p_{_j}\,u_{_{\{\kappa\}}}
\left(p_{_j}\right)\right]\,\frac{d\,p_{_j}}{d\,E_{_i}}+\frac{d\,\gamma}{d\,E_{_i}}\\
\nonumber
&=&\lambda\sum_ju_{_{\{\kappa\}}}\left(\frac{p(x_{_j})}{\alpha}\right)\,
\frac{d\,p(x_{_j})}{d\,x_{_j}}\,\frac{d\,x_{_j}}{d\,E_{_i}}
+\frac{d\,\gamma}{d\,E_{_i}}\\
\nonumber
&=&-\sum_jp_{_j}\,\left(\beta\,\delta_{_{ij}}+\frac{d\,\gamma}{d\,E_{_i}}\right)
+\frac{d\,\gamma}{d\,E_{_i}}\\
\nonumber&=&-\beta\,p_{_i} \ .\label{prob}
\end{eqnarray}
Thus, if the partition function is known, we can compute
the distribution function as
\begin{equation}
p_{_i}=-{1\over\beta}\,\frac{d}{d\,
E_{_i}}\,\ln_{_{\{\kappa\}}}\left(Z_{_\kappa}\right) \ .
\end{equation}
All the mean values of the macroscopical quantities
associated with the system can be expressed employing the
function $Z_{_\kappa}$:
\begin{equation}
\langle
A\,\rangle=\sum_ip_{_i}\,A_{_i}=-{1\over\beta}\,\sum_iA_{_i}\,\frac{d}{d\,
E_{_i}}\,\ln_{_{\{\kappa\}}}\left(Z_{_\kappa}\right) \
.\label{mean1}
\end{equation}
We recall that the microsate energies depend on other
macroscopical variables like the volume or the number of
particles. By following the standard literature
\cite{Callen}, let us introduce the canonically conjugate
variables through the relation
\begin{equation}
A_{_i}=-\frac{d\,E_{_i}}{d\,{\cal A}} \ .
\end{equation}
Then, Eq. (\ref{mean1}) assumes the expression
\begin{equation}
\langle A\,\rangle={1\over\beta}\,\frac{d}{d\, {\cal
A}}\,\ln_{_{\{\kappa\}}}\left(Z_{_\kappa}\right) \
,\label{mean2}
\end{equation}
which, taking into account the definition of $\kappa$-free
energy (\ref{f}), becomes
\begin{equation}
\langle A\,\rangle=-\frac{d\,F_{_\kappa}}{d\,{\cal A}} \
.\label{mv1}
\end{equation}
Let us observe that if the energies $E_{_i}$ depend on a
parameter $\varepsilon$ as
\begin{equation}
E_{_i}=E_{_i}^{(0)}+\varepsilon\,A_{_i} \ ,
\end{equation}
then the mean value of $A$ is given by
\begin{eqnarray}
\langle
A\,\rangle&=&-{1\over\beta}\,\frac{d}{d\,\varepsilon}
\,\ln_{_{\{\kappa\}}}\Big(Z_{_\kappa}(\beta,\,\varepsilon)\Big)
\ ,\label{mv2}
\end{eqnarray}
or equivalently
\begin{eqnarray}
\langle A\,\rangle&=&\frac{d}{d\,\varepsilon}
\,F_{_\kappa}(\beta,\,\varepsilon) \ .
\end{eqnarray}
This provides us with an alternative, useful trick for
calculating the expected values of an observable. In fact,
by adding the eigenvalues $A_{_i}$ of the observable $A$ to
the energy levels $E_{_i}$ we can calculate the new
partition function $Z_{_\kappa}(\beta,\,\varepsilon)$ and
the mean value $\langle A \rangle$ according to Eq.
(\ref{mv2}), and then, we set $\varepsilon$ to zero in the
final expression. Remark that this is analogous to the
source field method used in the path integral formulation
of quantum field theory.

\section{Conclusions}
In the framework of the generalized statistical mechanics
based on $\kappa$-entropy we have derived the canonical
partition function $Z_{_\kappa}(\beta)$ and studied its
main properties. This function plays a relevant role in the
formulation of $\kappa$-deformed thermostatistics theory
based on the entropy $S_{_\kappa}(U)$. It has been shown
that most of the thermodynamic variables of the system,
like the total energy $U$, the free energy
$F_{_\kappa}(1/\beta)$ and the entropy $S_{_\kappa}(U)$ can
be expressed in terms of the partition function
$Z_{_\kappa}(\beta)$ or its derivatives. We have shown an
algorithm to derive, starting from the expression of the
partition function, the mean value of all macroscopical
observables associated to the system.\\



\begin{thebibliography}{99}


\bibitem{Abe} S.~Abe, ``\textit{Nonextensive
Statistical Mechanics and its Applications}'', Y.~Okamoto
editors, (Springer, 2001).


\bibitem{Kaniadakis0} Special issue of Physica A \textbf{305},
Nos. 1/2 (2002), edited by G.~Kaniadakis, M.~Lissia, and
A.~Rapisarda.


\bibitem{Kaniadakis00} Special issue of Physica A \textbf{340}
Nos. 1/3 (2004), edited by G.~Kaniadakis, and M.~Lissia.


\bibitem{Kaniadakis1} G.~Kaniadakis, Physica\ A\ \textbf{296} (2001),
405.


\bibitem{Kaniadakis2}  G.~Kaniadakis, Phys.\ Lett.\ A\ \textbf{288} (2001), 282.


\bibitem{Kaniadakis3} G.~Kaniadakis, Phys.\ Rev.\ E\ \textbf{66} (2002), 056125.


\bibitem{Kaniadakis4} G.~Kaniadakis, Phys.\ Rev.\ E\ \textbf{72} (2005), 0261xy.


\bibitem{Scarfone1} G.~Kaniadakis, M.~Lissia, and
A.M.~Scarfone, Phys.\ Rev.\ E\ \textbf{71} (2005), 046128.


\bibitem{Scarfone2} G.~Kaniadakis, and A.M.~Scarfone,
Physica A \textbf{340} (2004), 102.


\bibitem{Scarfone3} G.~Kaniadakis, and A.M.~Scarfone, Physica A \textbf{305} (2002), 69.


\bibitem{Scarfone-Wada} A.M.~Scarfone, and T.~Wada, Phys.\ Rev.\ E (2005), in press.


\bibitem{Callen} H.B.~Callen, \textit{ Thermodynamics and an
Introduction to Thermostatistics} (Wiley, New York, 1985).


\bibitem{Greiner} W.~Greiner, L.~Neiser, and H.~St\"ocker,  \textit{Thermodynamics and
Statistical Mechanics} (Springer, New York, 1994).



\end{thebibliography}
\end{document}